# Report
## Genetic Testing for Complex Diseases: a Simulation Study Perspective

Nguyen Xuan Vinh,
Faculty of Information Technology, Monash University, VIC, Australia
Email: vinh.nguyen@monash.edu

It is widely recognized nowadays that complex diseases are caused by, amongst the others, multiple genetic factors. The recent advent of genome-wide association study (GWA) has triggered a wave of research aimed at discovering genetic factors underlying common complex diseases. While the number of reported susceptible genetic variants is increasing steadily, the application of such findings into diseases prognosis for the general population is still unclear, and there are doubts about whether the size of the contribution by such factors is significant. In this respect, some recent simulation-based studies have shed more light to the prospect of genetic tests. In this report, we discuss several aspects of simulation-based studies: their parameters, their assumptions, and the information they provide.

## 1. Introduction

It is widely recognized nowadays that complex diseases are caused by, amongst the others, multiple genetic factors.[1-4] The recent advent of genome-wide association study (GWA) has triggered a wave of research aimed at discovering genetic factors underlying common complex diseases. While the number of reported susceptible genetic variants is increasing steadily, the application of such findings into diseases prognosis for the general population is still unclear, and there are doubts about whether the size of the contribution by such factors is significant.[5] In this respect, simulation-based studies such as that conducted by Yang et. al.,[6] and more recently by Janssens et. al.,[7] are interesting and welcome contributions. While the former study showed that multiple genetic susceptibility tests (composed of up to 5 genes) significantly improve the prediction of multifactorial diseases, the latter was a more comprehensive study showing the prognostic accuracy of genetic tests composed of up to 400 genes, with various different scenarios for the risk allele frequencies and genotypes' odd ratios. In this report, we discuss several aspects of simulation-based studies, such as their model parameters, their assumptions, the information that they provide and more importantly, how to design a proper simulation methodology.

## 2. Background and Methods

As a common practice in the medical research literature, in this study we employ a Bayesian approach calculate the posterior probability of developing the disease given the genetic test results. Let $D$ and $D_o$ represent the diseased and non-diseased population respectively. Let $p = P(D)$ be the prior probability of developing the disease, also known as the disease prevalence or background risk, which tells us how likely a person chosen at random from the population is developing the disease, if no further information was available.

Let $G=\{g_1,g_2,...,g_n\}$ be the vector of test results for $n$ disease susceptibility genes. Now in the light of further evidences, i.e. the genetic test results $G$, the Bayes' rule can be employed to estimate the posterior risk of developing the disease:

$$P(D|G) = \frac{P(D)P(G|D)}{P(G)} \quad (1)$$

Now let $LR(G)$ denote the ratio of any observed $G$, defined as:

$$LR(G) = \frac{P(G|D)}{P(G|D_o)} \quad (2)$$

then it can be shown that the posterior probability of developing the disease given a particular genetic test results is:

$$P(D|G) = \frac{LR(G)P(D)}{[1-P(D)] + LR(G)P(D)} \quad (3)$$

Thus in order to calculate the post-test probability of developing the disease, we need to know the likelihood ratio of the genetic profile $G$.

### 2.1 Likelihood ratio estimation

If the $n$ genes are independent then:
$$P(G|D)=P(g_1|D)\,P(g_2|D)\ldots P(g_n|D)$$
and similarly:
$$P(G|D_o)=P(g_1|D_o)\,P(g_2|D_o)\ldots P(g_n|D_o).$$
It follows that:
$$LR(G)=LR(g_1)\,LR(g_2)\ldots LR(g_n) \quad (4)$$

Thus the likelihood ratio of a panel of independent test is simply the product of the likelihood ratios of the individual test results. As we shall see later, the likelihood ratios of the genotypes for each gene can be theoretically inferred, provided that information such as risk allele frequency, disease prevalence and odd ratios (or relative risk) of the risk genotypes is available. In practice, when such information is not always available, Yang et. al.[6] proposed a logistic regression based method in conjunction with a case-control study to estimate the likelihood ratios. More specifically, they showed that if a case-control study is designed such that there are $N_{CO}$ subjects in the control group and $N_{CA}$ subjects in the case group, then:

$$\ln LR(G) = \ln\left(\frac{N_{CO}}{N_{CA}}\right) + \alpha_{CC} + \beta G^T \quad (5)$$

where $\alpha_{CC}$ and $\beta$ are the intercept term and logistic regression coefficients of the odds of developing the disease respectively.

### 2.2 Performance evaluation

We consider first a single binary genetic test $G=\{g_1\}$ of which the value can take on only two values: 0 (negative) corresponding to the case of the gene is a homozygous non-risk genotype, and 1 (positive) otherwise. Various quality measures for binary test are presented in Table 1.

Table 1: Quality measures for a binary test.

| Single-gene test outcome | Disease status | | |
|---|---|---|---|
| | D (Diseased) | D$_o$ (Non-diseased) | |
| Positive | TP – True positive | FP – False Positive (Type I error, P-value) | Positive Predictive Value: PPV = TP/(FP+TP) |
| Negative | FN - False Negative (Type II Error) | TN – True Negative | Negative Predictive Value: NPV = TN/(FN+TN) |
| | Sensitivity = TP/(TP+FN) | Specificity = TN/(FP+TN) | |

For the case of multiple-gene tests $G=\{g_1,g_2,...,g_n\}$, if for each gene the test can take on two values, either negative or positive as considered by Yang *et. al.*,[6] then there would be $2^n$ different test results in the population. If a more detailed genetic model is assumed as by Janssens *et. al.*,[7] where the test result for each gene can take on three different values corresponding to the homozygous non-risk, heterozygous and homozygous risk genotypes, then there are $3^n$ different test results in the population. For these cases, in order to predict the disease status of a subject, one need to binarize its test result, i.e. make a final decision whether it is positive or negative. The decision can be made based upon choosing a suitable cut-off value for the posterior probability $P(D|G)$. Each cut-off value results in a different value of the test *sensitivity* and *specificity*, and there is often trade-off between these two quality measures. The discriminative accuracy of a multiple-gene test can be assessed by the area under receiver operating characteristic (ROC) - the plot of the *sensitivity* vs. (1 - *specificity*) for a test as its cut-off value is varied.

### 3. Simulation models
#### 3.1 A simple simulation model
In this section we reconsider the simulation study carried out by Yang *et. al.*[6] In this study the test result for each gene can take on two different values: 0 corresponding to the case of homozygous non-risk genotype, and 1 otherwise. All genes were assumed to be independent. To estimate the likelihood ratio of a Genetic test profile $G$, Yang *et. al.* employed a logistic regression model (as describe above), with a further assumption that genes interact multiplicatively. They reported the simulation results for genetics test composed of up to 5 genes, which included the likelihood ratio obtained from a simulated case-control study along with the 95% confidence interval. While estimating the likelihood ratio from a case-control study is certainly useful in practice, in our opinion, in a simulation study where information such as disease prevalence $p$, risk allele frequency $f$, and relative risk $R$ of risk genotypes, are available and controllable, it is more interesting to theoretically calculate the likelihood ratios. Likelihood ratio for each gene can be calculated based on Table 2.

Table 2: Basic table for calculating likelihood ratio of a binary single-gene test

| Single-gene test outcome | Disease status | | Total |
|---|---|---|---|
| | D (Diseased) | D$_o$ (Non-diseased) | |
| 1 | A | B | fN |
| 0 | C | D | (1-f)N |
| Total | pN | (1-p)N | N |

$N$: number of subjects in the study

The number of subjects in each category can be logically derived from the model parameters $f$, $p$ and $R$. More specifically, A, B, C, D is determined from the below equation system:

$$\begin{cases} A + B = fN \\ C + D = (1-f)N \\ A + C = Np \\ \dfrac{A/fN}{C/((1-f)N)} = R \end{cases} \quad (6)$$

of which an analytical solution can easily derived, and then the likelihood ratio of the test results can be calculated as:

$$LR(1) = \frac{A/pN}{B/((1-p)N)} \quad (7)$$

and

$$LR(0) = \frac{C/pN}{D/((1-p)N)} \quad (8)$$

The combined likelihood ratio of a multiple-gene test outcome can then be calculated by multiplying the likelihood ratios of its component test results. The likelihood ratio obtained is not an estimate but a single "true" value. Reporting this value seems to be more relevant than the value obtained from a simulated case-control study on a simulated population, since reflects the *intrinsic* ability of a multiple-gene test in discriminating the diseased group from the non-diseased group.

Based on our analysis, we re-perform the experiment corresponding to Table 1 in Yang *et. al.*'s paper. We consider genetic tests composed of up to 5 genes with relative risk ranging from 1.5 to 3.5, and gene frequency from 5% to 25%. Results are reported in Table 3. It can be observed that the estimated likelihood ratios deviate more from the "true" likelihood ratios when the number of gene in the model increases. In fact, our experiments (data not shown) have showed that when the number of subjects in the simulated case-control study is increased, *e.g* from 1000 to 10,000 subjects, then the estimated LR value gets very close to the true LR value. This is to be expected, since the accuracy of an unbiased estimator increases as the sample size increases. The estimated value might be lower or higher than the true value, reflecting the fact that it is not trivial to exactly estimate population parameters from a limited number of samples. We stress again that it is the "true" LR value which decides the theoretical *intrinsic* discriminative accuracy of a genetic test, and that why this value is of interest.

Table 3: Likelihood ratios and posterior probability of developing disease. Background risk $p = 5\%$.

| Genetic tests | Relative risk of genotype | Gene frequency | Estimated LR (95% CI)* | "True" LR | Posterior probability of developing disease (%) |
|---|---|---|---|---|---|
| One-gene tests: | | | | | |
| $g_1$ | 1.5 | 25 | 1.72 (1.37-2.16) | 1.36 | 6.7 |
| $g_2$ | 2.0 | 20 | 1.61 (1.22-2.13) | 1.72 | 8.3 |
| $g_3$ | 2.5 | 15 | 2.10 (1.55-2.85) | 2.16 | 10.2 |
| $g_4$ | 3.0 | 10 | 2.75 (1.88-4.03) | 2.71 | 12.5 |
| $g_5$ | 3.5 | 5 | 3.72 (2.21-6.26) | 3.50 | 15.6 |
| Two-gene tests (selected examples): | | | | | |
| $g_1$ | 1.5 | 25 | 2.95 (2.07-4.20) | 2.34 | 11 |
| $g_2$ | 2.0 | 20 | | | |
| $g_1$ | 1.5 | 25 | 3.6 (2.49-5.21) | 2.93 | 13.4 |
| $g_3$ | 2.5 | 15 | | | |
| $g_2$ | 2.0 | 20 | 3.26 (2.2-4.84) | 3.72 | 16.4 |
| $g_3$ | 2.5 | 15 | | | |
| Three-gene tests (selected examples): | | | | | |
| $g_1$ | 1.5 | 25 | 6.02 (3.79-9.56) | 5.05 | 21 |
| $g_2$ | 2.0 | 20 | | | |
| $g_3$ | 2.5 | 15 | | | |
| Four-gene tests (selected examples): | | | | | |
| $g_1$ | 1.5 | 25 | 19.0 (10.5-35.6) | 13.69 | 41.9 |
| $g_2$ | 2.0 | 20 | | | |
| $g_3$ | 2.5 | 15 | | | |
| $g_4$ | 3.0 | 10 | | | |
| Five-gene tests: | | | | | |
| $g_1$ | 1.5 | 25 | 151.7 (64.9-354.8) | 47.9 | 71.6 |
| $g_2$ | 2.0 | 20 | | | |
| $g_3$ | 2.5 | 15 | | | |
| $g_4$ | 3.0 | 10 | | | |
| $g_5$ | 3.5 | 5 | | | |

* results reported in Yang et. al, [6] estimated through a single simulated case-control study on a simulated population of $N$=1,000,000 subjects.

### 3.2 A more detailed simulation model

Janssens et al.[7] considered a more detailed genetic model where each gene has two alleles and three genotypes, i.e. the homozygous non-risk genotype ee, the homozygous risk genotype EE, and the heterozygous risk genotype Ee. Their study showed that it is theoretically possible to construct effective prognosis mechanisms based solely upon genetic profiling. The discriminative accuracy, evaluated by the area under the receiver-operating curve (AUC), can reach as high as 0.90 or more, using a test composed of multiple susceptibility genes, each with modest elevated relative risk. This simulation method has been employed in several subsequent studies.[1, 2, 8] In one of our current osteoporosis studies, a similar simulation strategy is being carried out to determine the contribution of genetic factors in the prognosis of bone fracture, on top of other clinical factors such as age, bone mineral density, history of fractures and falls. During the course of constructing the simulation study, we have proposed some changes and improvements that, in our opinion, may contribute both to the theoretical soundness and practical efficiency of the simulation methodology.

We first provide here a brief review of the original simulation procedure proposed by Janssens et al.[7], composed of three steps. The simulation model admits the following main parameters, namely the number of subjects in the studies, $N$, often set to 100,000; the number of genes in the study, $M$; the disease prevalence $p$; the allele frequencies $f=\{f_1, f_2,..., f_M\}$ where $f_i$ corresponds to the $i$-th gene; and the Odds Ratio $OR = \{OR_1, OR_2,..., OR_M\}$ of the heterozygous genotype corresponding to each gene.

**Janssens et al. simulation procedure for complex diseases prediction using multiple genes[7]:**

*Step 1 - Modeling genetic profiles*: Genetic profiles are built, consisting of up to $M = 400$ genes. Assuming that each single gene has two alleles and that all genotype and allele proportion were in Hardy-Weinberg equilibrium, the number of each genotype in the population can be calculated based on the parameter $f$. For each gene, a vector of length $N$ is created, containing the genotypes corresponding to $N$ subjects under study, with proportion of each genotype as specified by the parameter $f$. Each subject is assigned a genotype by randomly sampling (without replacement) this vector.

*Step 2 – Modeling disease risk associated with genetic profiles*: The disease risk associated with the genetics profiles is calculated using the Bayes' theorem:

$Posterior\_odds = LR\_of\_Genetic\_profile \times prior\_odds$

where $prior\_odds = p/(1-p)$ with $p$ being the disease prevalence parameter. Assuming no interaction between genes, the Likelihood Ratio (LR) of a genetic profile can be obtained by multiplying the LRs of the single genotypes:

$$LR\_of\_Genetic\_profile = \prod_{i=1}^{M} LR(Genotype_i)$$

The LR of a single genotype is the percentage of the genotype among subjects who will develop the disease divided by the percentage of the genotype among subjects who will not develop the disease. The LR is calculated from Table 4.

**Table 4:** Number of genotypes of the $i$-th gene in the population

| Genotype | Subject with disease | Subject without disease | Total |
|---|---|---|---|
| $EE_i$ | A | B | $f_i^2 N$ |
| $Ee_i$ | C | D | $2f_i(1-f_i)N$ |
| $ee_i$ | E | G | $(1-f_i)^2 N$ |
| Total | $Np$ | $N(1-p)$ | $N$ |

The Odds Ratio of the heterozygous risk genotype $Ee_i$ of the $i$-th gene, is given as a model parameter, namely $OR_i$, while the Odds Ratio of the homozygous risk genotype $EE_i$ is assumed to be the square of $OR_i$. In order to calculate the LRs, the value of $A$-$G$ must be determined through solving the following equation system:

$$\begin{cases} A + B = f_i^2 N \\ C + D = 2f(1-f_i)N \\ E + G = (1-f_i)^2 N \\ A + C + E = Np \\ AG/(BE) = OR_i^2 \\ CG/(DE) = OR_i \end{cases} \quad (9)$$

Then the LR of each genotype for the $i$-th gene can be calculated as: $LR(EE_i)=A(1-p)/(Bp)$; $LR(Ee_i)=C(1-p)/(Dp)$ and $LR(ee_i)=E(1-p)/(Gp)$.

*Step 3- Modeling disease status*: The disease risk of each subject is compared with a randomly drawn value between 0 and 1 from a uniform distribution. A subject is assigned to the group who will develop the disease when the disease risk is higher than the random value and to the group who will not develop the disease otherwise. This procedure ensures that subjects with high disease risks are more likely to be assigned to the group who will develop the disease than those with low risks.

After all the three steps have been completed, we have in hand a simulated population with known disease status and known risk profile. The data can then be statistically analysed, for example, by calculating the AUC of a prognosis test built upon the genetic profile using the Bayesian risk model described above. During the course of our experiment with the simulation procedure described above, we noticed several points that might be further improved.

The first proposed improvement involves the solution of equation (9). Despite its modest appearance, it is rather difficult to solve this equation system analytically. In the absence of a closed form solution, Janssens *et al.* (personal communications) resorted to a heuristic iterative algorithm briefly described as follows: starting with $A=1$, with knowledge of the genotype frequencies, the value for $B$ can be inferred. Table 4 then essentially reduces to a 2x2 table of the 4 unknown variables $C$-$G$ (similar to Table 2), for which the values can be filled in analytically. The odds ratio of $AG/BE$ is then checked; if it is too low then $A$ is increased by one. The process is repeated until $AG/BE$ becomes close enough to $(OR_i)^2$. While this algorithm seems to work in practice, an analytical solution for (9) is still of great interest, both for the theoretical soundness of the simulation methodology and for its efficient implementation, since the number of repeated experiments is often huge.

Indeed, we are now showing that such a closed form solution is achievable. Let $t_1$, $t_2$, $t_3$, $t_4$, $t_5$ denote the values of $f_i^2 N$, $2f_i(1-f_i)N$, $(1-f_i)^2 N$, $pN$ and $OR_i$ respectively, by careful algebraic manipulation it is possible to show that the value of $D$ is given by a suitable solution of a third order polynomial equation of the form:

$$A_3 D^3 + A_2 D^2 + A_1 D + A_0 = 0$$

where the coefficients are given by:
$A_3 = 1 - 2t_5 + t_5^2$
$A_2 = -t_1 t_5^2 - 2t_5 t_4 + t_4 + t_5 t_1 + t_3 t_5 - t_3 + 4t_5 t_2 - 2t_2 + t_4 t_5^2 - 2t_2 t_5^2$
$A_1 = t_3 t_2 - 2t_1 t_5 t_2 + 2t_4 t_5 t_2 + t_1 t_5^2 t_2 - 2t_5 t_3 t_2 - t_4 t_2 + t_2^2 - t_4 t_5^2 t_2 - 3t_5 t_2^2 + t_2^2 t_5^2$
$A_0 = -t_5 t_4 t_2^2 + t_5 t_1 t_2^2 + t_5 t_3 t_2^2 + t_5 t_2^3$

Since a closed form solution for third order polynomial equation exists, finding the closed form solution for $D$ from this equation, and subsequently the suitable values for $A$-$G$ is rather straightforward. A short program in the R software for this purpose is available from us upon request.

The second major proposed modification in the simulation procedure involves the following observation: the proportion of subjects developing the disease in the obtained simulated population, denoted by $p'$, is often smaller than the disease prevalence as prescribed by the parameter $p$, and the gap grows as the number of genes in the study gets larger. For example, let $p=0.3$, then when the number of genes $M=100$, $p'$ is only ~0.26 and this figure reduces to ~0.22 when $M=200$ and only ~0.17 when $M=400$. Since $p \neq p'$, a question arises as to what the real meaning of the parameter $p$ is. Since $p$ represents the disease prevalence in the population, it is probably preferable that the proportion $p'$ of subjects developing the disease in the simulated population is kept close to $p$ or ideally exactly equal to $p$. So where does the inconsistency come from? Pondering this question, we turn to the following, more basic conceptual question of cause and consequence: is the disease risk value (as predicted by a prognosis model such as the Bayesian model described above) the cause or the consequence of the disease status? The answer is probably clear: the disease risk value is the consequence of the disease status. In the simulation procedure employed by Janssens *et al.*,[7] this cause-consequence relationship has been reversed, i.e., in *step 3*, the disease status is the consequence of the disease risk value predicted by the Bayesian model. If another risk model, such as the logistic model, were employed simultaneously for comparison, then possibly the disease status of a subject derived from the two models would be different. This is again another inconsistency since the value of a predictive test should only give a prediction about the disease status, but not modify the "true" disease status of a subject.

To rectify the above stated inconsistencies, it is necessary to make a major change in the conceptual design of the simulation method, that is, to put the objects in the right place in the cause-consequence relationship. In light of our analysis above, we propose a new simulation procedure as follows (compare with the above procedure by Janssens *et al.*[7]):

*Step 1- Modeling disease status*: create two groups of subjects containing correspondingly *Np* subjects who is developing the disease, and *N(1-p)* subjects who will is not.

*Step 2 - Modeling genetic profiles*: for each gene, solve equation (9) to fill in Table 4. Now for each group we can create a genetic pool where the number of each genotype is prescribed by Table 4. Next, assign each subject a random genotype by randomly sampling, without replacement, the pool corresponds to its group. Repeat the same process for all genes in the simulation.

*Step 3 – Modeling disease risk associated with genetic profiles*: in this final step, a predictive test based on the genetic profile is built. Examples are the above describe Bayesian model, or a logistic model.

With this new simulation procedure, the simulated population obtained now closely reflects the intended population as prescribed by the set of model parameters. We repeated the simulation corresponding to Figure 2 in the paper by Janssens *et al.*[7] In particular, the discriminative accuracy (in terms of the area under the ROC) for multiple-gene tests consisted of one to 400 genes was calculated. The risk allele frequency for all genes was set to 10%, and the ORs for all genes was the same. Our simulation results, presented in Figure 1, showed a slight difference from that reported by Janssens *et al.*[7], especially when the number of genes in the simulation is higher, where the difference of *p* and *p'* in Janssens *et al.*'s simulation procedure is large. More specifically, our AUC value obtained at such values of *M* is slightly higher. Though the difference seems to be minor in this particular scenario, we nevertheless advocate the use of our simulation model for its consistency and implementation efficiency.

Using the likelihood ratios calculated from Table 4 it is also able to answer another interesting question: how many genes are needed for a PPV value five times greater than the background risk? We assume that all genes contain a risk genotype, *i.e.* being either *EE* or *Ee*, and have the same OR for the heterozygous risk genotype (the OR for the homozygous risk genotype is assumed to be the square of the OR for the heterozygous risk genotype). Clearly, the smallest number of genes needed for a particular PPV is achieved when all genotypes in the test are *EE*'s, and the most number is required when all genotypes are *Ee*'s. Results are reported in Table 5.

**Table 5:** Number of genes needed for a PPV value 5 times greater than the background risk *p* which is set to 10%. OR's for all genes are the same. All genes are assumed to bear risk allele.

| OR | Number of genes needed | |
|---|---|---|
| | Smallest | Largest |
| 1.05 | 29 | 76 |
| 1.10 | 15 | 40 |
| 1.25 | 7 | 18 |
| 1.5 | 4 | 10 |
| 3.0 | 2 | 5 |
| 5.0 | 1 | 4 |

In summary, in this report we have discussed several aspects of simulation based studies for assessing the prospect of genetic testing. We discussed several previous simulation-based studies and proposed simulation procedure, which is more efficient and consistent. Using this procedure, it is also possible to assessed the effectiveness of genetic test on top of other classical clinical factor and this is one of our ongoing work.

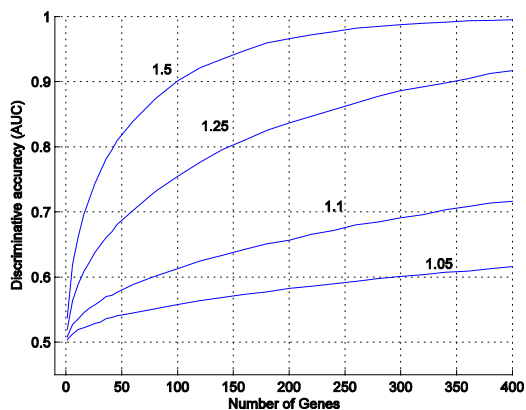

**Figure 1.** Discriminative accuracy, evaluated by the area under the ROC (AUC), of multiple-gene tests. Disease prevalence *p*=10%, risk allele frequency is 10% for all genes. The numbers next to each line represents the OR for the heterozygous genotype of all genes included in the test..